\documentclass[aps,prd,floats,floatfix,nofootinbib
]{revtex4-1}
\usepackage{graphicx,amsmath}
\usepackage{amssymb}
\usepackage{amsfonts}
\usepackage{subfig}

\begin{document}

\title{Constraints on large scalar multiplets from perturbative unitarity}

\author{Katy Hally}
\email{khally@physics.carleton.ca}

\author{Heather E.~Logan}
\email{logan@physics.carleton.ca}

\author{Terry Pilkington}
\email{tpilking@physics.carleton.ca}

\affiliation{Ottawa-Carleton Institute for Physics, Carleton University, Ottawa, Ontario K1S 5B6, Canada}

\date{March 5, 2012}

\begin{abstract}
We determine the constraints on the isospin and hypercharge of a scalar electroweak multiplet from partial-wave unitarity of tree-level scattering diagrams.  The constraint from SU(2)$_L$ interactions yields $T \leq 7/2$ (i.e., $n \leq 8$) for a complex scalar multiplet and $T \leq 4$ (i.e., $n \leq 9$) for a real scalar multiplet, where $n = 2T+1$ is the number of isospin states in the multiplet.  
\end{abstract}

\maketitle

\section{Introduction}

Extensions of the scalar sector of the Standard Model (SM) beyond the usual single SU(2)$_L$-doublet Higgs field are, as yet, largely unconstrained by experiment.  Such extensions are common in models that address the hierarchy problem of the SM, such as supersymmetric models~\cite{SUSY} and little Higgs models~\cite{LH}, as well as in models for neutrino masses, dark matter, etc.  Most of these extensions contain additional SU(2)$_L$-singlet, -doublet, and/or -triplet scalar fields.  

Some extensions of the SM contain scalars in larger multiplets of SU(2)$_L$.  Such larger multiplets have been used to produce a natural dark matter candidate~\cite{Cirelli:2005uq}, which is kept stable thanks to an accidental global U(1) symmetry present in the Higgs potential for multiplets with $T \geq 2$.  Three different models with a Higgs quadruplet (isospin $T=3/2$) have also been proposed for neutrino mass generation~\cite{Babu:2009aq,Picek:2009is,Ren:2011mh}.  Scalar multiplets larger than doublets with significant vacuum expectation values have long been known to run afoul of the tight experimental constraints on the $\rho$ parameter~\cite{PDG2010}; however, the tree-level relation $\rho \equiv M_W^2/M_Z^2 \cos^2\theta_W = 1$ is automatically satisfied for multiplets that obey the relation~\cite{rhomults,HHG}
\begin{equation}
	(2T+1)^2 - 3 Y^2 = 1,
	\label{eq:rho}
\end{equation}
where $T$ is the isospin of the multiplet and $Y$ is the hypercharge, related to the electric charge by $Q = T^3 + Y/2$.  This condition is satisfied by an unlimited number of $(T,Y)$ combinations.  The smallest few are $T=1/2$, $Y=1$ (the usual SM Higgs doublet); $T=3$, $Y=4$ (a 7-plet containing a maximally charged state $\chi^{+5}$); $T=25/2$, $Y = 15$ (a 26-plet containing a maximally charged state $\chi^{+20}$); etc.  Other multiplets are allowed if their vacuum expectation values are small or zero, or if a cancellation of their contributions to $\rho$ is arranged using custodial SU(2) symmetry~\cite{GM} or fine-tuning.

In this paper we aim to constrain the proliferation of large scalar multiplets using perturbative unitarity of tree-level scattering amplitudes.  Perturbative unitarity of tree-level scattering amplitudes has most famously been used to set an upper limit on the mass of a weakly-coupled Higgs boson~\cite{LQT}.  The bounds coming from perturbative unitarity can be violated at the cost of making the theory strongly coupled.  In our case, a scalar multiplet with a large weak charge has correspondingly large $2 \to 2$ tree-level scattering amplitudes for scalar pair annihilation into electroweak gauge bosons.  Requiring that the zeroth partial wave amplitudes remain smaller than the unitarity bound constrains the maximum isospin and hypercharge of a large scalar multiplet.  Larger multiplets would violate the unitarity bound at tree-level; in this case higher-order corrections to the scattering amplitude must restore unitarity, implying that the weak sector has become strongly coupled.  

In what follows we compute the $2 \to 2$ scattering amplitudes for scalar pair annihilation into electroweak gauge bosons, for arbitrary values of the isospin and hypercharge of the scalar multiplet.  We perform the coupled channel analysis including all relevant initial and final states.  Imposing the unitarity bound, we show that tree-level perturbative unitarity constrains a complex scalar SU(2)$_L$ multiplet to have isospin $T \leq 7/2$, and a real scalar SU(2)$_L$ multiplet to have $T \leq 4$.\footnote{For comparison, we note that Ref.~\cite{Cirelli:2005uq} quotes an upper bound of $T \leq 3$ for a real scalar multiplet at the TeV scale, derived by assuming that the scalar multiplet is the only addition to the theory beyond the SM and requiring that its contribution to the renormalization group running of the SU(2)$_L$ gauge coupling does not drive this coupling nonperturbative below the Planck scale.  From the same requirement we find an upper bound of $T \leq 5/2$ for a complex scalar multiplet.  Our limit from tree-level unitarity is less constraining but more generally applicable.}  We also set corresponding limits on the hypercharge.

This paper is organized as follows.  In Sec.~\ref{sec:couplings} we present the scattering amplitudes for a generic scalar SU(2)$_L$ multiplet scattering into electroweak gauge bosons.  In Sec.~\ref{sec:coupchan} we perform the coupled channel analysis, derive general expressions for the largest amplitude eigenvalues, and apply the unitarity constraint.  Finally, in Sec.~\ref{sec:conclusions} we discuss the implications of our results and conclude.  Details of the matrix element calculations are given in the Appendix.

\section{Couplings and matrix elements}
\label{sec:couplings}

To obtain the desired unitarity constraints, we study scattering of two scalars into two electroweak gauge bosons in the high-energy limit, for overall electrically-neutral initial and final states.  We are interested in the constraints that arise from large electroweak charges; therefore we ignore electroweak symmetry breaking and work in the unmixed ${\rm SU}(2)_L \times {\rm U}(1)_Y$ basis.  This has the advantage of allowing us to cleanly separate the constraints due to the SU(2)$_L$ and U(1)$_Y$ interactions.  We also thus consider only the transverse polarization states of the gauge bosons and ignore the gauge boson masses.  

The gauge interactions of the scalars arise from the scalar gauge-kinetic terms,
\begin{equation}
	\mathcal{L} \supset \left\{ \begin{array}{ll}
	 (\mathcal{D}_{\mu} X)^{\dagger} (\mathcal{D}^{\mu} X) & {\rm for} \ X \ {\rm complex}, \\
	 \frac{1}{2} (\mathcal{D}_{\mu} \Xi)^{\dagger} (\mathcal{D}^{\mu} \Xi) & {\rm for} \ \Xi \ {\rm real}.
	 \end{array} \right.
\end{equation}
We will express the complex and real scalar multiplets in the charge basis as 
\begin{equation}
	X = \left( \begin{array}{c} \chi_1 \\ \chi_2 \\ \vdots \\ \chi_n \end{array} \right),
	\qquad
	\Xi = \left( \begin{array}{c} \xi^Q \\ \vdots \\ \xi^0 \\ \vdots \\ \xi^{-Q} \end{array} \right).
\end{equation} 
Note that for the real multiplet, $Y$ must be zero and $T$ must be an integer.  Note also that $\xi^0$ is a real scalar, while the neutral member of $X$ (if one exists) is a complex scalar.  The positively and negatively charged states in $\Xi$ are related by $(\xi^{Q})^* = (-1)^Q \xi^{-Q}$.  For $X$ we also have $T^3 \chi_1 = T \chi_1$, $T^3 \chi_n = -T \chi_n$, etc., where $T$ is the total isospin of the multiplet $X$ and $T^3$ is the third component of the isospin.  

The covariant derivative is given as usual by
\begin{eqnarray}
	\mathcal{D}_{\mu} &=& \partial_{\mu} - i g W^a_{\mu} T^a - i g^{\prime} B_{\mu} \frac{Y}{2}  
	\nonumber \\
	&=& \partial_{\mu} - i \frac{g}{\sqrt{2}} \left( W^+_{\mu} T^+ + W^-_{\mu} T^- \right) 
	- i g W^3_{\mu} T^3 - i g^{\prime} B_{\mu} \frac{Y}{2},
\end{eqnarray}
where $T^a$ are the SU(2) generators and $W^{\pm}$and $T^{\pm}$  are given by
\begin{eqnarray}
	W^{\pm}_{\mu} &=& \frac{1}{\sqrt{2}} \left( W^1_{\mu} \mp i W^2_{\mu} \right), \nonumber \\
	T^{\pm} &=&  T^1 \pm i T^2.
	\label{eq:generators}
\end{eqnarray}

The partial wave amplitudes are related to scattering matrix elements according to
\begin{equation}
	\mathcal{M} = 16 \pi \sum_J (2J + 1) a_J P_J(\cos\theta),
\end{equation}
where $J$ is the orbital angular momentum of the final state and $P_J(\cos\theta)$ is the corresponding Legendre polynomial.  Tree-level partial wave unitarity dictates that 
\begin{equation}
	|{\rm Re}\, a_0| \leq 1/2.
	\label{eq:ubound}
\end{equation}  
We will use only the zeroth partial wave amplitude, $a_0$, to set our unitarity limits.  

\begin{figure}
\centering
\subfloat[]{\label{fig:four-point}\includegraphics[scale=0.55]{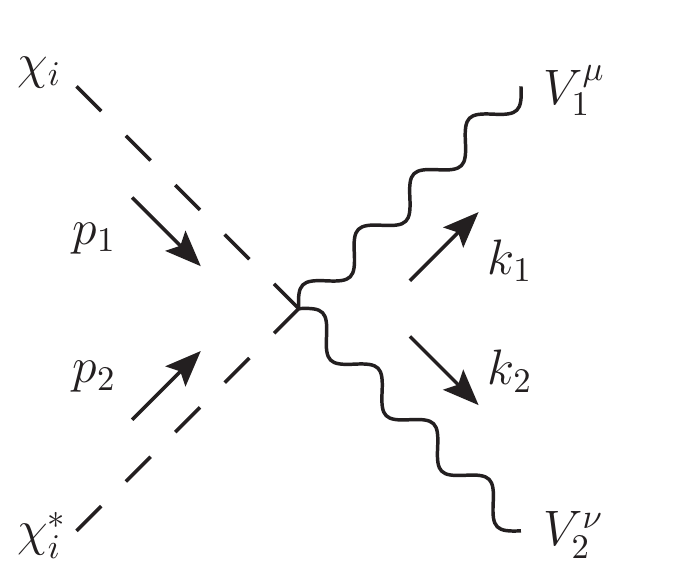}}\hspace*{1em}
\subfloat[]{\label{fig:t-channel}\includegraphics[scale=0.55]{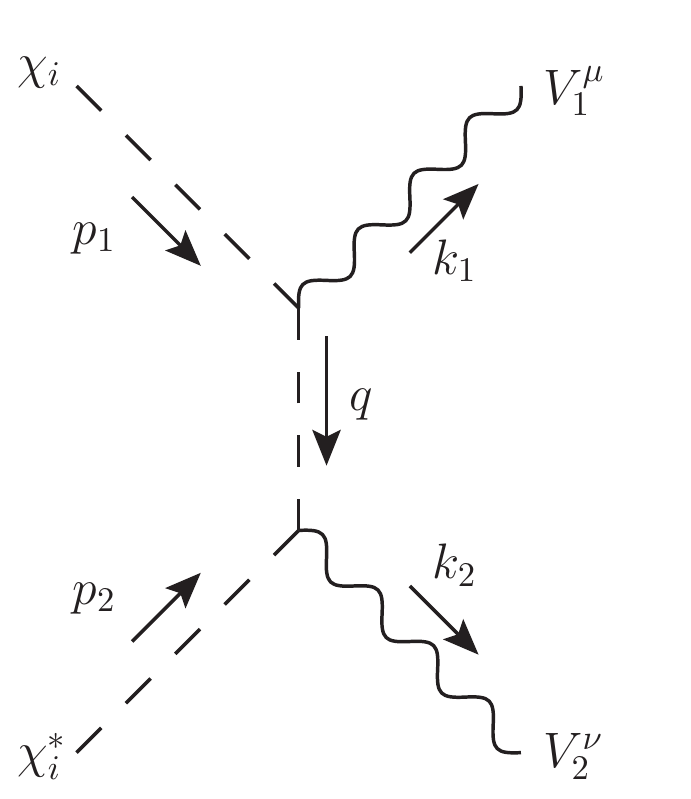}}\hspace*{1em} 
\subfloat[]{\label{fig:u-channel}\includegraphics[scale=0.55]{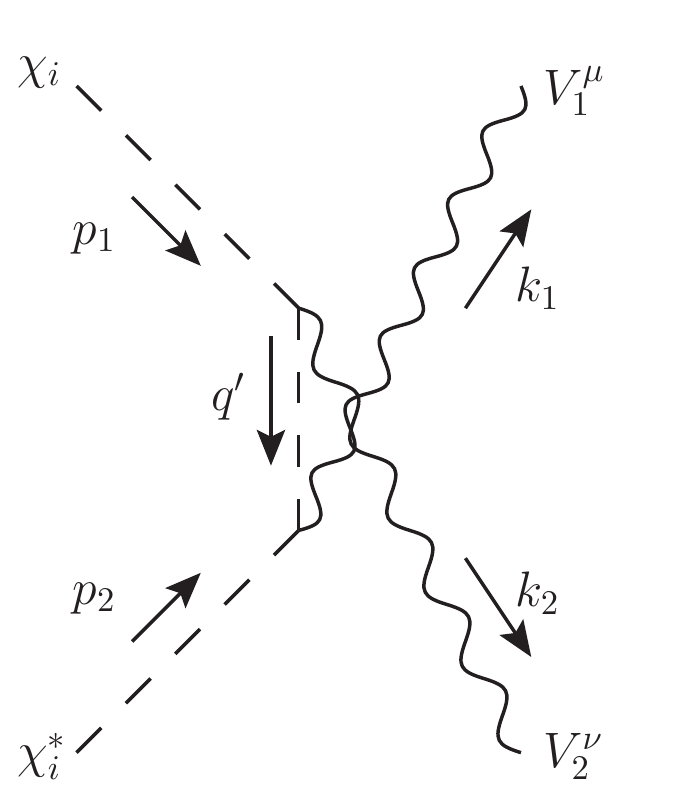}}
\subfloat[]{\label{fig:s-channel}\includegraphics[scale=0.55]{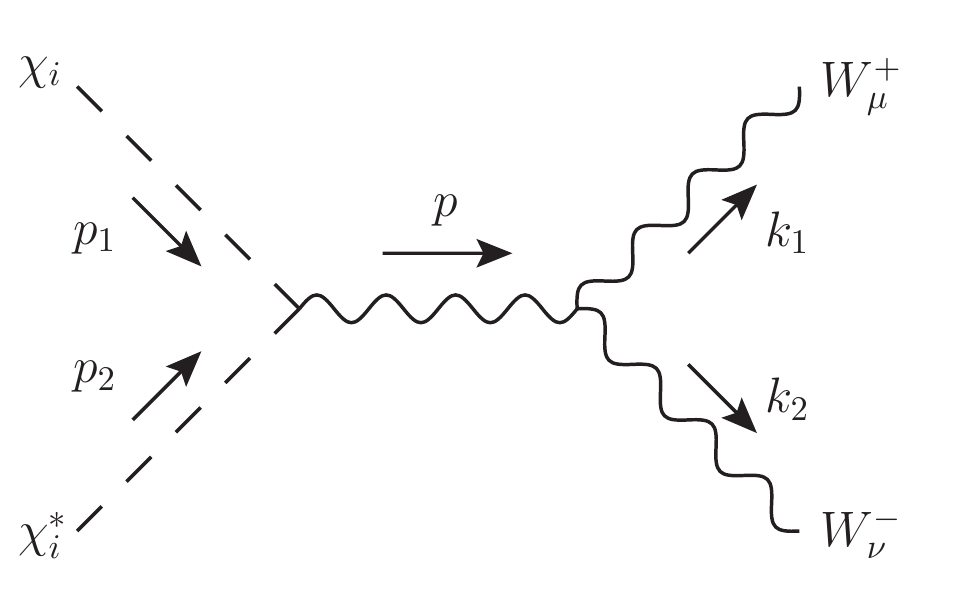}}
\caption{Feynman diagrams contributing to $\chi \chi^* \to V_1V_2$.}
\label{fig:fd}
\end{figure}

The contributing Feynman diagrams are shown in Fig.~\ref{fig:fd}.  Diagrams (a), (b), and (c) contribute to the processes $\chi^* \chi \to BB$, $W^3W^3$, and $BW^3$, while all four diagrams contribute to the process $\chi^* \chi \to W^+W^-$.  
The matrix elements are computed in the Appendix.  For each final state, there are four distinct polarization combinations of the gauge bosons; two combinations give zero for the matrix element, while the other two each yield the same zeroth partial wave matrix element in the high-energy limit.

For the complex scalar $X$ we find,
\begin{eqnarray}
	a_0(\chi_i^* \chi_i \to BB/\sqrt{2}) &=& \frac{g^2}{16\pi} \frac{s^2_W}{c^2_W} \frac{Y^2}{2\sqrt{2}},
	\nonumber \\
	a_0(\chi_i^* \chi_i \to BW^3) &=& \frac{g^2}{16\pi} \frac{s_W}{c_W} T^3 Y,
	\nonumber \\
	a_0(\chi_i^* \chi_i \to W^3W^3/\sqrt{2}) &=& \frac{g^2}{16\pi} \sqrt{2} (T^3)^2,
	\nonumber \\
	a_0(\chi_i^* \chi_i \to W^+W^-) &=& \frac{g^2}{16\pi} \left[ T(T+1) - (T^3)^2 \right],
	\label{eq:a0complex}
\end{eqnarray}
where $s_W$ ($c_W$) is the sine (cosine) of the weak mixing angle defined via $g^{\prime}/g = s_W/c_W$, and we have used the fact that initial or final states involving two identical particles receive an extra $1/\sqrt{2}$ normalization.  

For the real scalar $\Xi$ we find, 
\begin{eqnarray}
	a_0(\xi^{Q*} \xi^Q \to W^3W^3/\sqrt{2}) &=& \frac{g^2}{16\pi} \sqrt{2} (T^3)^2 \qquad {\rm for} \ T^3 = Q \geq 1,
	\nonumber \\
	a_0(\xi^{Q*} \xi^Q \to W^+W^-) &=& \frac{g^2}{16\pi} \left[ T(T+1) - (T^3)^2 \right] \qquad {\rm for} \ T^3 = Q \geq 1,
	\nonumber \\
	a_0(\xi^0\xi^0/\sqrt{2} \to W^3W^3/\sqrt{2}) &=& 0,
	\nonumber \\
	a_0(\xi^0\xi^0/\sqrt{2} \to W^+W^-) &=& \frac{g^2}{16\pi} \frac{1}{\sqrt{2}} T(T+1).
\end{eqnarray}
Note that the main difference between the real and complex scalars is in the multiplicity of scalar states.

\section{Coupled channel analysis}
\label{sec:coupchan}

When nonzero amplitudes exist that couple the same initial (final) state to multiple final (initial) states, the strongest unitarity bound comes from applying Eq.~(\ref{eq:ubound}) to the largest eigenvalue of the matrix of amplitudes of all the channels thus coupled.

\subsection{U(1)$_Y$ interactions}

From Eq.~(\ref{eq:a0complex}) we see that the zeroth partial wave amplitude for $\chi_i^* \chi_i \to BB/\sqrt{2}$ is the same for all $n$ members of the multiplet $X$.  Including only U(1)$_Y$ interactions, the coupled-channel matrix in the basis $(BB/\sqrt{2}, \chi_1^* \chi_1, \ldots, \chi_n^* \chi_n)$ is thus given by
\begin{equation}
	a_0 = \sqrt{2} \frac{g^2}{16\pi} \frac{s_W^2}{c_W^2} \frac{Y^2}{2\sqrt{2}}
	\left( \begin{array}{cccc}
	0 & 1 & \cdots & 1 \\
	1 & 0 & \cdots & 0 \\
	\vdots & \vdots &  & \vdots \\
	1 & 0 & \cdots & 0 \end{array} \right),
\end{equation}
where the $\sqrt{2}$ in front comes from the two contributing gauge boson polarization combinations.
The matrix of integers in the preceding equation has a pair of nonzero eigenvalues, $\sqrt{n}$ and $-\sqrt{n}$, as well as $n-1$ zero eigenvalues.  The eigenvectors corresponding to the nonzero eigenvalues are 
\begin{equation}
	\frac{1}{\sqrt{2}} \left[ (BB/\sqrt{2}) \pm (\chi^*\chi)_{\rm sym} \right],
\end{equation}
where we define the properly-normalized symmetric combination of all $n$ states $\chi_i^*\chi_i$ (i.e., the combination with zero total isospin) according to
\begin{equation}
	(\chi^*\chi)_{\rm sym} \equiv \frac{1}{\sqrt{n}} \sum_i \chi_i^* \chi_i.
	\label{eq:chisym}
\end{equation}
The nonzero eigenvalues of the zeroth partial wave amplitude matrix involving only U(1)$_Y$ interactions are therefore given by $\pm a_0^{\rm max, U(1)}$, where
\begin{equation}
	a_0^{\rm max,U(1)} =  \frac{g^2}{16\pi} \frac{s_W^2}{c_W^2} \frac{Y^2}{2} \sqrt{n}.
	\label{eq:a0maxU1}
\end{equation}
Imposing the unitarity bound in Eq.~(\ref{eq:ubound}) and plugging in numbers,\footnote{We use $\alpha_{\rm em} = s^2_W g^2/4\pi \simeq 1/128$ and $s^2_W \simeq 0.231$.  These values are valid at the weak scale; logarithmic renormalization-group running of $g$ and $g^{\prime}$ will cause numerical variations in our results at higher mass scales.} we obtain a constraint on the hypercharge as a function of the size of the multiplet,
\begin{equation}
	|Y| \lesssim \frac{19.8}{n^{1/4}}.
\end{equation}

Note that when more than one hypercharged scalar multiplet is present, the largest eigenvalue of the coupled-channel matrix is just the sum in quadrature of the largest eigenvalue for each multiplet $X_j$:
\begin{equation}
	a_0^{\rm max,U(1)} = \left[ \sum_j \left( a_0^{{\rm max,U(1)},j} \right)^2 \right]^{1/2}.
\end{equation}

\subsection{SU(2)$_L$ interactions}

We first consider the complex multiplet $X$.  The coupled channel analysis for the SU(2)$_L$ interactions is complicated by the fact that the scattering amplitude is not the same for all $n$ initial states $\chi_i^*\chi_i$.  The coupled channel matrix in the basis $(W^+W^-, W^3W^3/\sqrt{2}, \chi_1^*\chi_1, \ldots \chi_n^*\chi_n)$ is given by
\begin{equation}
	a_0 = \sqrt{2} \frac{g^2}{16\pi} \left( \begin{array}{ccccc}
	0 & 0 & A_1 & \cdots & A_n \\
	0 & 0 & B_1 & \cdots & B_n \\
	A_1 & B_1 & 0 & \cdots & 0 \\
	\vdots & \vdots & \vdots & & \vdots \\
	A_n & B_n & 0 & \cdots & 0 \end{array} \right),
	\label{eq:su2matrix}
\end{equation}
where again the $\sqrt{2}$ in front comes from the two contributing gauge boson polarization combinations, and
\begin{eqnarray}
	A_i &=& T(T+1) - (T^3)^2, \nonumber \\
	B_i &=& \sqrt{2} (T^3)^2,
\end{eqnarray}
with $T^3$ evaluated for the appropriate state $\chi_i^* \chi_i$.  

The matrix in Eq.~(\ref{eq:su2matrix}) has two pairs of nonzero eigenvalues, together with $n-2$ zero eigenvalues.  The first (and largest) pair is $\pm a_0^{{\rm max,SU}(2)}$, where
\begin{equation}
	a_0^{{\rm max,SU}(2)} = \frac{g^2}{16\pi} 2 T (T+1) \sqrt{\frac{n}{3}}  
	= \frac{g^2}{16\pi} \frac{(n^2 - 1) \sqrt{n}}{2\sqrt{3}}.
	\label{eq:a0maxSU2}
\end{equation}
The eigenvectors corresponding to the first pair of eigenvalues are
\begin{equation}
	\frac{1}{\sqrt{2}} \left[ (WW)_{\rm sym} \pm (\chi^*\chi)_{\rm sym} \right],
\end{equation}
where $(\chi^*\chi)_{\rm sym}$ is given in Eq.~(\ref{eq:chisym}) and $(WW)_{\rm sym}$ is the symmetric (isospin zero) combination of the SU(2)$_L$ gauge fields given by
\begin{equation}
	(WW)_{\rm sym} = \frac{1}{\sqrt{3}} \left[ \sqrt{2} (W^+W^-) + (W^3W^3/\sqrt{2}) \right]
	= \frac{1}{\sqrt{6}} \left[ W^1W^1 + W^2W^2 + W^3W^3 \right].
\end{equation}
A similar analysis for a real multiplet $\Xi$ yields
\begin{equation}
	a_0^{{\rm max,SU}(2)}({\rm real}) = \frac{1}{\sqrt{2}} a_0^{{\rm max,SU}(2)}({\rm complex})
	= \frac{g^2}{16\pi} \frac{(n^2 - 1) \sqrt{n}}{2\sqrt{6}},
	\label{eq:a0maxreal}
\end{equation}
where for a real multiplet $n$ must be an odd integer.  As in the U(1)$_Y$ case, when more than one scalar multiplet carrying isospin is present, the largest eigenvalue of the coupled-channel matrix is the sum in quadrature of the largest eigenvalue [Eq.~(\ref{eq:a0maxSU2}) or (\ref{eq:a0maxreal})] for each multiplet.

Imposing the unitarity bound in Eq.~(\ref{eq:ubound}) upon the largest eigenvalue [Eqs.~(\ref{eq:a0maxSU2}) and (\ref{eq:a0maxreal})] and plugging in numbers, we obtain an upper bound on the size (or isospin) of a scalar multiplet from tree-level perturbative unitarity of the SU(2)$_L$ interaction alone:
\begin{eqnarray}
	n \leq 8 &\quad (T \leq 7/2)& \qquad {\rm for \ a \ complex \ multiplet}, \nonumber \\
	n \leq 9 &\quad (T \leq 4)& \qquad {\rm for \ a \ real \ multiplet}.
	\label{eq:nboundSU2}
\end{eqnarray}

For completeness we give here the second (smaller) pair of nonzero eigenvalues.  These are $\pm a_0^{\perp}$, where\footnote{In the second equality we used $\sum_{j=1}^N j^2 = N(N+1)(2N+1)/6$ and $\sum_{j=1}^N j^4 = N(N+1)(2N+1)(3N^2+3N-1)/30$.}
\begin{equation}
	a_0^{\perp} = \frac{g^2}{16\pi} \sqrt{\frac{2}{3}} \left[ \sum_i \left[ T(T+1) - 3(T^3)^2 \right]^2 \right]^{1/2}
	= \frac{g^2}{16\pi} \frac{\sqrt{n(n^2-1)(n^2-4)}}{\sqrt{30}},
\end{equation}
where again we have included the extra factor of $\sqrt{2}$ coming from the two contributing gauge boson polarization combinations.  The eigenvectors corresponding to these eigenvalues are
\begin{equation}
	\frac{1}{\sqrt{2}} \left[ (WW)_{\perp} \pm (\chi^*\chi)_{\perp} \right],
\end{equation}
where 
\begin{eqnarray}
	(WW)_{\perp} &=& \frac{1}{\sqrt{3}} \left[ (W^+W^-) - \sqrt{2} (W^3W^3/\sqrt{2}) \right],
	\nonumber \\
	(\chi^*\chi)_{\perp} &=& \frac{\sum_i \chi_i^* [T(T+1) - 3 (T^3)^2] \chi_i}{\left[n(n^2-1)(n^2-4)/20 \right]^{1/2}}.
\end{eqnarray}
These are the neutral components of the combinations with total isospin 2 and are orthogonal to $(WW)_{\rm sym}$ and $(\chi^*\chi)_{\rm sym}$, respectively.

\subsection{Combined electroweak gauge interactions}

Complex scalar multiplets that carry both isospin and hypercharge couple the SU(2)$_L$ and U(1)$_Y$ channels together, as well as introducing the additional $BW^3$ channel.  

First, we observe that the $BW^3$ channel is not coupled to the other channels.  This is because, after diagonalization of the coupled channel matrix, $BW^3$ couples to the linear combination of scalars,
\begin{equation}
	(\chi^*\chi)_{T^3} = \frac{\sum_i \chi_i^* T^3 \chi_i}{\left[ \sum_i (T^3)^2 \right]^{1/2}}
	= \frac{\sum_i \chi_i^* T^3 \chi_i}{\left[ n (n^2 - 1) / 12 \right]^{1/2}},
\end{equation}
where the sums run over the $n$ members $\chi_i$ of the multiplet.  This linear combination of scalars is the neutral component of the combination with total isospin 1 and is orthogonal to $(\chi^*\chi)_{\rm sym}$ and $(\chi^*\chi)_{\perp}$.  The pair of nonzero eigenvalues corresponding to the $BW^3$ channel are $\pm a_0^{BW^3}$, where
\begin{equation}
	a_0^{BW^3} = \frac{g^2}{16\pi} \frac{s_W}{c_W} \frac{Y \sqrt{n(n^2 - 1)}}{\sqrt{6}},
\end{equation}
where again we have included the extra factor of $\sqrt{2}$ coming from the two contributing gauge boson polarization combinations.
For any values of $Y$ and $n$, this eigenvalue is always smaller than the one we will find in Eq.~(\ref{eq:a0coupled}) below.

Second, we recall that the combination of scalars $(\chi^*\chi)_{\perp}$ that couples to $(WW)_{\perp}$ is orthogonal to $(\chi^*\chi)_{\rm sym}$; therefore it does not couple to the $(BB/\sqrt{2})$ channel.  The corresponding eigenvalue $a_0^{\perp}$ is always smaller than the one corresponding to $(WW)_{\rm sym}$, so it is not of interest to us.

Finally, we observe that $(BB/\sqrt{2})$ and $(WW)_{\rm sym}$ both couple to the same linear combination of scalars, i.e., $(\chi^*\chi)_{\rm sym}$.  The corresponding eigenvalue, which is the largest eigenvalue of the full coupled-channel system, is then obtained by adding in quadrature the corresponding eigenvalues for the U(1)$_Y$ and SU(2)$_L$ couplings:\footnote{Or equivalently, by diagonalizing the matrix
\begin{equation}
	a_0^{\rm max,sym} = \left( \begin{array}{ccc}
	0 & 0 & a_0^{{\rm max,U}(1)} \\
	0 & 0 & a_0^{{\rm max,SU}(2)} \\
	a_0^{{\rm max,U}(1)} & a_0^{{\rm max,SU}(2)} & 0
	\end{array} \right),
\end{equation}
in the basis $[(BB/\sqrt{2}), (WW)_{\rm sym}, (\chi^*\chi)_{\rm sym}]$.}
\begin{equation}
	a_0^{\rm max, sym} = \left[ \left( a_0^{{\rm max, U}(1)} \right)^2 + \left( a_0^{{\rm max, SU}(2)} \right)^2 \right]^{1/2},
	\label{eq:a0coupled}
\end{equation}
where $a_0^{{\rm max, U}(1)}$ and $a_0^{{\rm max, SU}(2)}$ are the eigenvalues given in Eqs.~(\ref{eq:a0maxU1}) and (\ref{eq:a0maxSU2}), respectively.  

This upper bound is most relevant for scalar multiplets that carry both isospin and hypercharge.  
In Table~\ref{tab:Y} we give the upper limit on the hypercharge $Y$ allowed by perturbative unitarity for a single complex scalar multiplet with isospin $T$.  Note in particular that in all cases a multiplet with $Y = 2T$ is allowed; in such a multiplet the state $\chi_n$ with $T^3 = -T$ is electrically neutral.  Note also that the multiplet with $T=3$, $Y=4$, which can have a nonzero vacuum expectation value while preserving $\rho = 1$ at tree level, is allowed.

\begin{table}
\begin{tabular}{c cccc cccc}
\hline\hline
$n$ & 1 & 2 & 3 & 4 & 5 & 6 & 7 & 8 \\
$T$ & 0 & 1/2 & 1 & 3/2 & 2 & 5/2 & 3 & 7/2 \\
$|Y_{\rm max}|$ & 19.8 & 16.7 & 15.1 & 14.0 & 13.0 & 12.1 & 10.8 & 8.3 \\
\hline\hline
\end{tabular}
\caption{Upper limit on the hypercharge $Y$ allowed by perturbative unitarity for a complex $n$-plet of SU(2)$_L$.}
\label{tab:Y}
\end{table}

We finally note that, when more than one scalar multiplet is present, the largest eigenvalue of the coupled-channel matrix can be found efficiently as follows.  First, the maximum eigenvalues for the U(1)$_Y$ and SU(2)$_L$ interactions can be computed separately for each multiplet using Eqs.~(\ref{eq:a0maxU1}) and (\ref{eq:a0maxSU2}).  Then the largest eigenvalue of the full coupled-channel system is just the largest eigenvalue of the following matrix, 
\begin{equation}
	a_0^{\rm max,sym} = \left( \begin{array}{ccccc}
	0 & 0 & a_0^{{\rm max, U}(1), 1} & \cdots & a_0^{{\rm max, U}(1), N} \\
	0 & 0 & a_0^{{\rm max, SU}(2), 1} & \cdots & a_0^{{\rm max, SU}(2), N} \\
	a_0^{{\rm max, U}(1), 1} & a_0^{{\rm max, SU}(2), 1} & 0 & \cdots & 0 \\
	\vdots & \vdots & \vdots & & \vdots \\
	a_0^{{\rm max, U}(1), N} & a_0^{{\rm max, SU}(2), N} & 0 & \cdots & 0 
	\end{array} \right),
	\label{eq:multiple}
\end{equation}
where we work in the basis $[(BB/\sqrt{2})$,  $(WW)_{\rm sym}$, $(\chi^*\chi)_{{\rm sym}, 1}, \ldots, (\chi^*\chi)_{{\rm sym}, N}]$
and the index $1, \ldots, N$ counts the scalar multiplets.

\section{Conclusions}
\label{sec:conclusions}

In this paper we have derived upper limits on the isospin and hypercharge of a complex or real scalar transforming under SU(2)$_L \times {\rm U}(1)_Y$ by requiring that tree-level scattering amplitudes for two scalars annihilating into two electroweak gauge bosons satisfy the unitarity bound.  Violation of this condition implies that the weak gauge sector becomes strongly coupled at energies above the scalar's mass.  Our main results are the expressions for the largest eigenvalue of the coupled-channel scattering amplitude matrix [Eq.~(\ref{eq:a0maxU1}) for hypercharge and Eqs.~(\ref{eq:a0maxSU2}) and (\ref{eq:a0maxreal}) for SU(2)$_L$] and the procedure for combining the amplitudes from multiple scalars [Eq.~(\ref{eq:multiple})].

We find that the perturbative unitarity bound is satisfied for a complex scalar multiplet with $T \leq 7/2$ (i.e., $n \leq 8$) or a real scalar multiplet with $T \leq 4$ (i.e., $n \leq 9$; recall that real multiplets must have integer $T$).  In particular, of the multiplets whose vacuum expectation values preserve $\rho = 1$ at tree level [see Eq.~(\ref{eq:rho})], only the SM doublet and the complex scalar with $T = 3$, $Y = 4$ are allowed in a weakly coupled theory; larger representations violate perturbative unitarity.

The constraints become more stringent if more than one large multiplet is present.  For example, 
perturbative unitarity of the SU(2)$_L$ interactions allows only one complex 8-plet ($T = 7/2$).  Similarly, perturbative unitarity allows two complex 7-plets, but adding a third violates perturbative unitarity; in particular, this implies that a color-triplet 7-plet is forbidden if SU(2)$_L$ is to remain weakly coupled.  Finally, a real color-octet scalar must have $T \leq 2$ in order to preserve perturbative unitarity of SU(2)$_L$.

\appendix
\section{Calculation of scattering amplitudes}

For concreteness we define the scattering process in the $x$--$z$ plane, with momenta
\begin{eqnarray}
	p_1^{\mu} &=& (E_{p_1}, \ |\vec p_1| \sin\theta, \ 0, \ |\vec p_1| \cos\theta), \nonumber \\
	p_2^{\mu} &=& (E_{p_2}, \ -|\vec p_2| \sin\theta, \ 0, \ -|\vec p_2| \cos\theta), \nonumber \\
	k_1^{\mu} &=& (E_{k_1}, \ 0, \ 0, \ |\vec k_1|), \nonumber \\
	k_2^{\mu} &=& (E_{k_2}, \ 0, \ 0, \ -|\vec k_2|),
\end{eqnarray}
where $p_1$ and $p_2$ are the incoming four-momenta of $\chi_i$ and $\chi_i^*$, $k_1$ and $k_2$ are the outgoing four-momenta of $V_1$ and $V_2$, respectively, and $\theta$ is the scattering angle.  We also define transverse polarization basis vectors for the gauge bosons according to
\begin{eqnarray}
	\epsilon_{\rm out}^{\mu}(k_1) &=& (0, \ 0, \ 1, \ 0), \qquad
	\epsilon_{\rm out}^{\mu}(k_2) = (0, \ 0, \ -1, \ 0), \nonumber \\
	\epsilon_{\rm in}^{\mu}(k_1) &=& (0, \ 1, \ 0, \ 0), \qquad
	\epsilon_{\rm in}^{\mu}(k_2) = (0, \ 1, \ 0, \ 0),
\end{eqnarray}
where the subscripts ``out'' and ``in'' refer to polarizations out of and in the scattering plane, respectively.  The signs are chosen for later convenience.

We first consider the process $\chi_i^* \chi_i \to B_{\mu} B_{\nu}$ for a state $\chi_i$ in a complex scalar multiplet.  The first three diagrams in Fig.~\ref{fig:fd} contribute.  The relevant couplings are given by
\begin{eqnarray}
	\chi_i \chi_i^* B_{\mu} B_{\nu} : &\qquad& i g^{\prime 2} \frac{Y^2}{2} g_{\mu\nu}, \nonumber \\
	\chi_i (q_1) \chi_i^* (q_2) B_{\mu} : &\qquad& -i g^{\prime} \frac{Y}{2} (q_1 - q_2)_{\mu},
\end{eqnarray}
with all particles and momenta incoming.  The matrix elements for the four-point, $t$-channel, and $u$-channel diagrams are
\begin{eqnarray}
	\mathcal{M}_a &=& g^{\prime 2} \frac{Y^2}{2} \epsilon^{\mu}(k_1) \epsilon_{\mu}(k_2), 
	\nonumber \\
	\mathcal{M}_b &=& -g^{\prime 2} \frac{Y^2}{4} (p_1 + q)^{\mu} \epsilon_{\mu}(k_1)
	(q - p_2)^{\nu} \epsilon_{\nu}(k_2) \frac{1}{q^2 - m_i^2}, \nonumber \\
	\mathcal{M}_c &=& -g^{\prime 2} \frac{Y^2}{4} (q^{\prime} - p_2)^{\mu} \epsilon_{\mu}(k_1)
	(p_1 + q^{\prime})^{\nu} \epsilon_{\nu}(k_2) \frac{1}{q^{\prime 2} - m_i^2},
\end{eqnarray}
where $q = p_1 - k_1 = k_2 - p_2$ and $q^{\prime} = p_1 - k_2 = k_1 - p_2$ are the $t$- and $u$-channel momenta, respectively, and $m_i$ is the mass of $\chi_i$.

We now evaluate these matrix elements for the four transverse polarization combinations of the gauge bosons.  For both gauge bosons polarized out of the scattering plane we have $\mathcal{M}_b = \mathcal{M}_c = 0$ and
\begin{equation}
	\mathcal{M}_a = \mathcal{M}_{\rm tot} = g^{\prime 2} \frac{Y^2}{2}.
\end{equation}
When one gauge boson is polarized out of the scattering plane and the other is polarized in the plane, all three diagrams give zero.  Finally, when both gauge bosons are polarized in the scattering plane, we have
\begin{eqnarray}
	\mathcal{M}_a &=& -g^{\prime 2} \frac{Y^2}{2}, \nonumber \\
	\mathcal{M}_b &=& -g^{\prime 2} Y^2 \frac{|\vec p_1| |\vec p_2| \sin^2\theta}{q^2 - m_i^2},
	\nonumber \\
	\mathcal{M}_c &=& -g^{\prime 2} Y^2 
	\frac{|\vec p_1| |\vec p_2| \sin^2\theta}{q^{\prime 2} - m_i^2}.
\end{eqnarray}
The second and third amplitudes simplify significantly in the high-energy limit.  Working in the center-of-mass frame we can substitute $|\vec p_1| = |\vec p_2| = \sqrt{s}/2$, 
$q^2 = t = - s (1 - \cos\theta)/2$, and $q^{\prime 2} = u = - s (1 + \cos\theta)/2$.  We can neglect the $m_i^2$ in the propagators without danger from the $t$- and $u$-channel singularities because the $\sin^2 \theta = (1 + \cos\theta)(1 - \cos\theta)$ in the numerator cancels the divergences in the dangerous regions of phase space.  In the high-energy limit we then obtain,
\begin{eqnarray}
	\mathcal{M}_b &=& g^{\prime 2} \frac{Y^2}{2} (1 + \cos\theta), \nonumber \\
	\mathcal{M}_c &=& g^{\prime 2} \frac{Y^2}{2} (1 - \cos\theta).
\end{eqnarray}
The total amplitude for both gauge bosons polarized in the scattering plane is then
\begin{equation}
	\mathcal{M}_{\rm tot} = g^{\prime 2} \frac{Y^2}{2},
\end{equation}
which is the same as that for both gauge bosons polarized out of the plane.

The matrix element calculations for $\chi_i^* \chi_i \to W^3_{\mu} W^3_{\nu}$ and $\chi_i^* \chi_i \to B_{\mu} W^3_{\nu}$ go through in exactly the same way, with the coupling replacements
\begin{eqnarray}
	\left( g^{\prime} \frac{Y}{2} \right)^2 &\rightarrow& \left( g T^3 \right)^2 
	\quad {\rm for} \ W^3 W^3, \nonumber \\
	\left( g^{\prime} \frac{Y}{2} \right)^2 &\rightarrow& \left( g^{\prime} \frac{Y}{2} \right)
	\left( g T^3 \right)
	\quad {\rm for} \ B W^3.
\end{eqnarray}

The matrix element calculation for $\chi_i^* \chi_i \to W^+_{\mu} W^-_{\nu}$ is more complicated due to the presence of the fourth diagram in Fig.~\ref{fig:fd} involving the $s$-channel exchange of $W^3$.  Furthermore, the scalars exchanged in the $t$- and $u$-channel diagrams have different masses in general.  The relevant couplings are,
\begin{eqnarray}
	\chi_i \chi_i^* W^-_{\mu} W^+_{\nu} : &\qquad& i \frac{g^2}{2} \left[ T^+ T^- + T^- T^+ \right] 
	g_{\mu\nu}, \nonumber \\
	\chi_i(q_1) \chi_{i+1}^*(q_2) W^-_{\mu} : &\qquad& -i \frac{g}{\sqrt{2}} T^- (q_1 - q_2)_{\mu},
	\nonumber \\
	\chi_i(q_1) \chi_{i-1}^*(q_2) W^+_{\mu} : &\qquad& -i \frac{g}{\sqrt{2}} T^+ (q_1 - q_2)_{\mu},
	\nonumber \\
	\chi_i(q_1) \chi_i^*(q_2) W^3_{\mu} : &\qquad& -i g T^3 (q_1 - q_2)_{\mu},
	\nonumber \\
	W^3_{\rho}(p) W^-_{\mu}(-k_1) W^+_{\nu}(-k_2) : &\qquad&
	i g \left[ g_{\mu\nu} (k_2 - k_1)_{\rho} + g_{\nu\rho} (-p - k_2)_{\mu} 
	+ g_{\rho\mu} (p + k_1)_{\nu} \right],
\end{eqnarray}
with all particles and momenta incoming.  Here $\chi_{i+1}$ ($\chi_{i-1}$) is the state with $T^3$ value one unit lower (higher) than $\chi_i$.  We write the couplings involving $W^{\pm}$ in terms of the generators $T^{\pm}$ for later convenience.  Note that from Eq.~(\ref{eq:generators}) we can write \begin{eqnarray}
	T^{\pm} T^{\mp} &=& (T^1 \pm i T^2)(T^1 \mp i T^2) 
	= T^1 T^1 + T^2 T^2 \mp i [T^1,T^2] \nonumber \\
	&=& (\vec T)^2 - (T^3)^2 \pm T^3 
	= T(T+1) - (T^3)^2 \pm T^3,
\end{eqnarray}
where we used the SU(2) commutation relation and applied the $(\vec T)^2$ operator.  From this we obtain $\left[ T^+T^- + T^- T^+ \right] = 2 \left[ T(T+1) - (T^3)^2 \right]$.

For both $W$ bosons polarized out of the scattering plane, we have $\mathcal{M}_b = \mathcal{M}_c = 0$ as before,
\begin{equation}
	\mathcal{M}_a = \frac{g^2}{2} \left[ T^+T^- + T^- T^+ \right]
	= g^2 \left[ T(T+1) - (T^3)^2 \right],
\end{equation}
and 
\begin{equation}
	\mathcal{M}_d = - g^2 T^3 \frac{1}{p^2} (p_1 - p_2)_{\rho} (k_2 - k_1)^{\rho},
	\label{eq:Md}
\end{equation}
where $p = p_1 + p_2 = k_1 + k_2$ is the $s$-channel four-momentum.
Here we have used the Feynman-gauge propagator for a massless gauge boson, $-i g^{\rho \sigma}/p^2$, for the $s$-channel $W^3$.  This is legitimate because we are working in the electroweak theory before electroweak symmetry breaking.  Ghosts do not contribute.  Working in the center-of-mass frame and taking the high-energy limit, the momentum dot product in Eq.~(\ref{eq:Md}) becomes
\begin{equation}
	(p_1 - p_2)_{\rho} (k_2 - k_1)^{\rho} = 4 \, \vec p_1 \cdot \vec k_1 = s \cos\theta.
\end{equation}
Thus the matrix element for the $s$-channel diagram is
\begin{equation}
	\mathcal{M}_d = -g^2 T^3 \cos\theta.
\end{equation}
This is proportional to the first Legendre polynomial $P_1(\cos\theta) = \cos\theta$ and thus contributes only to the first partial wave amplitude $a_1$.  Our result for the matrix element contributing to the zeroth partial wave, for both $W$ bosons polarized out of the scattering plane, is therefore
\begin{equation}
	\mathcal{M}_{{\rm tot}, 0} = g^2 \left[ T(T+1) - (T^3)^2 \right].
\end{equation}

When one $W$ boson is polarized out of the scattering plane and the other is polarized in the plane, all four diagrams give zero.  Finally, when both $W$ bosons are polarized in the scattering plane, we have in the high-energy limit,
\begin{eqnarray}
	\mathcal{M}_a &=& - \frac{g^2}{2} \left[ T^+ T^- + T^- T^+ \right], \nonumber \\
	\mathcal{M}_b &=& g^2 T^+ T^- (1 + \cos\theta), \nonumber \\
	\mathcal{M}_c &=& g^2 T^- T^+ (1 - \cos\theta), \nonumber \\
	\mathcal{M}_d &=& g^2 T^3 \cos\theta,
\end{eqnarray}
where we have followed the same steps as before to simplify the $t$-, $u$-, and $s$-channel diagrams.  Once again $\mathcal{M}_d$ contributes only to the first partial wave amplitude, as do the parts of $\mathcal{M}_b$ and $\mathcal{M}_c$ that are proportional to $\cos\theta$.  The angle-independent parts of the first three diagrams sum up to yield a matrix element contributing to the zeroth partial wave, for both $W$ bosons polarized in the scattering plane, of
\begin{equation}
	\mathcal{M}_{\rm tot, 0} = \frac{g^2}{2} \left[ T^+ T^- + T^- T^+ \right]
	= g^2 \left[ T(T+1) - (T^3)^2 \right].
\end{equation}
Once again, this is the same as the matrix element for both $W$ bosons polarized out of the scattering plane.

\begin{acknowledgments}
We thank B.~Coleppa for useful conversations and K.~Earl for collaboration during the early stages of this project.  H.E.L.\ thanks J.~Gunion and H.~Haber for early discussions that ultimately inspired this project.
This work was supported by the Natural Sciences and Engineering
Research Council of Canada.
\end{acknowledgments}


\end{document}